\documentclass[a4paper,11pt]{article}

\usepackage{jcappub} 
\usepackage{graphics,color}
\usepackage{bm}
\usepackage{graphicx}
\usepackage{amsmath}
\usepackage{amssymb}
\usepackage{enumitem}
\usepackage{tensor}	
\usepackage[fleqn]{mathtools}
\usepackage{epstopdf}
\usepackage{multirow}
\usepackage{booktabs}
\usepackage{xcolor}
\usepackage[margin=1in]{geometry}
\usepackage{array} 

\def\be{\begin{equation}}
\def\ee{\end{equation}}
\def\ba{\begin{eqnarray}}
\def\ea{\end{eqnarray}}
\def\l{\left}
\def\r{\right}
\def\f{\frac}

\def\eft{\texttt{EFTCAMB}}
\def\heft{$\mathcal{H}-$\texttt{EFTCAMB}}

\usepackage{graphicx}
\usepackage{color}
\usepackage{bm}
\usepackage{longtable}
\usepackage{amssymb}
\usepackage{rotating}
\usepackage{hyperref}

\title{\heft: A Cobaya-Integrated, Python-Wrapped Extension of EFTCAMB for Covariant Horndeski Gravity}

\author[1,2,3]{Gen Ye,}

\author[4,5]{Shijie Lin,}

\author[6]{Jiaming Pan,}

\author[1]{Dani de Boe,}

\author[1]{Stan Verhoeve,}

\author[7]{Marco Raveri,}

\author[4,5]{Bin Hu,}

\author[8,9]{Noemi Frusciante,}

\author[1]{Alessandra Silvestri}

\affiliation[1]{Institute Lorentz, Leiden University, PO Box 9506, Leiden 2300 RA, The Netherlands}

\affiliation[2]{Universit\'e de Gen\`eve, D\'epartement de Physique Th\'eorique and Centre for Astroparticle Physics, 24 quai Ernest-Ansermet, CH-1211 Gen\`eve 4, Switzerland}

\affiliation[3]{School of Physics, University of Chinese Academy of Sciences, Beijing 100049, China}

\affiliation[4]{Institute for Frontier in Astronomy and Astrophysics, Beijing Normal University, Beijing 102206, China}

\affiliation[5]{School of Physics and Astronomy, Beijing Normal University, Beijing 100875, China}

\affiliation[6]{Department of Physics and Leinweber Institute for Theoretical Physics,\\
 University of Michigan, 450 Church St, Ann Arbor, MI 48109}

\affiliation[7]{Department of Physics and INFN, University of Genova, Via Dodecaneso 33, 16146, Italy}

\affiliation[8]{Dipartimento di Fisica ``E. Pancini", Universit\`a degli Studi  di Napoli  ``Federico II", Compl. Univ. di Monte S. Angelo, Edificio G, Via Cinthia, I-80126, Napoli, Italy}

\affiliation[9]{ INFN Sezione di Napoli, Università degli Studi di Napoli “Federico II”, Compl. Univ. di Monte S. Angelo, Edificio G, Via Cinthia, I-80126, Napoli, Italy}

\emailAdd{Gen.Ye@unige.ch}
\emailAdd{linsj@mail.bnu.edu.cn}
\emailAdd{jiamingp@umich.edu}
\emailAdd{deboe@lorentz.leidenuniv.nl}
\emailAdd{verhoeve@strw.leidenuniv.nl}
\emailAdd{Marco.Raveri@unige.it}
\emailAdd{bhu@bnu.edu.cn}
\emailAdd{noemi.frusciante@unina.it}
\emailAdd{silvestri@lorentz.leidenuniv.nl}

\abstract{We present \heft, the official successor to \eft. The original \eft~is designed as a consistent and numerically stable implementation of the effective field theory (EFT) of dark energy in the Einstein-Boltzmann code \texttt{CAMB}. On top of this, \heft~introduces a new Horndeski module that supports computing cosmology for an arbitrary input covariant Horndeski Lagragian. \heft \ supports both mapping the Horndeski theory to an EFT lagrangian to solve in the EFT framework as well as directly solving for the scalar field equations of motion derived from the covariant Lagrangian. The latter approach also works for the cases when the Horndeski field experiences turn-overs, e.g. oscillation, where the EFT approach breaks down. The Horndeski module has been validated by comparing internally with existing models in the original \eft~and externally with \texttt{hi\_class}. \heft~features a flexible Python wrapper that is seamlessly integrated into the widely utilized cosmological sampler \texttt{Cobaya}. \heft~is publicly available and serves as a comprehensive tool for testing gravity against the precision data from current and next-generation surveys. \heft~is publicly available at \href{https://github.com/EFTCAMB/EFTCAMB}{this URL}.}

\begin{document}
\maketitle
\flushbottom

\section{Introduction} \label{sec:intro}

Modern observational cosmology has entered the era of precision measurements, driven by local distance ladder approaches~\cite{SNIa2016A,SNIa2022A,DistanceL2019F,DistanceL2025F}, a series of cosmic microwave background (CMB) experiments~\cite{WMAP,Planck2018}, and large-scale structure (LSS) surveys~\cite{SDSSBAO,DESY3,KiDs-Legacy,DESIDR2}. Within the framework of General Relativity (GR), the accelerating expansion of the Universe is typically attributed to a mysterious component dubbed dark energy. In the simplest scenario, dark energy is described by a cosmological constant $\Lambda$, leading to the well-known $\Lambda$CDM model.

Alternatively, the late-time accelerated expansion may arise from a modification of gravity on cosmological scales, rather than from a new energy component. These theories, collectively referred to as modified gravity (MG), propose that GR may not be the complete description of gravity at large distances. Testing such scenarios against observations requires numerical tools capable of evolving both the background and linear perturbations in the presence of dynamical dark energy or modified gravity.

To this end, several numerical tools based on the public Einstein–Boltzmann solvers \texttt{CAMB}~\cite{CAMB} and \texttt{CLASS}~\cite{CLASS} have been developed, such as \eft~\cite{EFTCAMB1,EFTCAMB2}, \texttt{MGCAMB}~\cite{NewMGCAMB,MGCAMB_ini} and \texttt{hi\_class}~\cite{hiCLASS,Bellini:2019syt}. 
A minimal and widely studied extension of GR introduces a single additional scalar degree of freedom. A unified and systematic framework for such theories is the effective field theory of dark energy 
(EFTofDE)~\cite{EFTofDE-BJ,EFTofDE-BJs,EFTofDE-GG,EFTofDE-GJ,EFTofDE-PF}, which encapsulates the linear dynamics of scalar perturbations around a Friedmann–Robertson–Walker (FRW) background using a small set of time-dependent functions.

Built on this formalism, \eft~\cite{EFTCAMB1,EFTCAMB2,NumericalNotes3} is a publicly available extension of \texttt{CAMB} that enables accurate and consistent calculations of cosmological observables in both model-independent EFT parameterizations and specific modified gravity theories. The framework describes the background evolution and linear perturbations in terms of a set of generic EFT operators, constructed as an expansion around a flat FRW background. Each operator captures a distinct physical effect, providing a unified description of a wide class of models. This model-independent approach allows for systematic and stable exploration of deviations from GR, while also enabling a mapping between EFT operators and underlying covariant theories. In this way, \eft~serves as an important bridge between fundamental theory and observational cosmology.

In this spirit, \eft~has played a significant role in the analysis of cosmological data over the past decade, serving as a key tool for both theoretical modeling and data interpretation. It has been used in analyses of \textit{Planck} data~\cite{PlanckVI,PlanckXIV}, joint constraints combining CMB and LSS measurements~\cite{EFT-DESI-fullshape}, and forecasting studies for next-generation surveys such as \textit{Euclid}~\cite{EuclidfR,EuclidLinear,EuclidParame}. More recently, analyses incorporating DESI BAO data~\cite{DESIDR1,DESIDR2} have suggested that the observed hint of a phantom crossing may indicate the presence of non-minimally coupled gravity within the \eft~framework~\cite{NMG-DR1,hqwq-m19h}. Together, these studies highlight the robustness of \eft~as a comprehensive framework for confronting modified gravity and dark energy models with current and future data.

Alongside these applications, many dark energy and modified gravity models have been mapped to the \eft~framework~\cite{EFTCAMB1,Horava,BeyondH,K-mouflage,ExtendedG,EFTquintessence,ScalingCG,ShiftS,CubicG-Ye,FreezingG}. As the theoretical landscape has expanded and observational constraints have become increasingly precise, there is a growing need for numerical tools capable of handling a broad spectrum of covariant, physically motivated theories. The most general scalar–tensor theory with at most second-order equations of motion in four-dimensional spacetime is the Horndeski theory~\cite{Horndeski1974,Horndeski50}, which has attracted significant attention due to its generality and phenomenological viability. It was shown in~\cite{Kobayashi:2011nu} that Horndeski theory is equivalent to the generalized Galileon theory~\cite{Deffayet_2009,Nicolis_2009,Deffayet_2011,Charmousis_2012}, obtained by covariantizing Galileon theories originally formulated in flat spacetime. Horndeski theory encompasses many of the viable dark energy and modified gravity models that are of interest for cosmological tests.

Motivated by these developments, in \heft~, the official successor to \eft, we introduce a dedicated module capable of handling any model within the Horndeski class directly from its covariant Lagrangian. The $\mathcal{H}$ in the name stands for covariant Horndeski. This development represents a major step toward bridging the gap between high-energy theory and cosmological phenomenology. In addition, the new implementation includes a fully non-parametric mode that allows users to specify arbitrary functional forms for the EFT functions, as well as for the covariant Horndeski Lagrangian functions. This feature enables robust, data-driven reconstructions of cosmic acceleration without assuming a specific underlying theory. Prior to this release, the development version of \heft~has already been used in several research papers~\cite{NMG-DR1,hqwq-m19h,Ye:2024kus,Ye:2024zpk,Yao:2025wlx}.

The new release is publicly available at \url{https://github.com/EFTCAMB/EFTCAMB}, together with a Python wrapper that seamlessly integrates into the widely utilized cosmological sampler \texttt{Cobaya} \cite{Torrado:2020dgo, 2019ascl.soft10019T}. In Section.\ref{sec:newEFT} we first briefly review the EFT formalism and the original \eft, then describe the major new features of \heft. We then present the validation of the new functionalities through internal consistency checks and comparisons with external codes in Section.\ref{sec:Compare}. Finally, we showcase the application of \heft~in Section.\ref{sec:tgs} by studying a specific modified gravity theory dubbed thawing gravity.

\section{Overview of \heft} \label{sec:newEFT}

The original public releases of \eft~\cite{EFTCAMB2,EFTCAMB1} provided a consistent and numerically stable implementation of the EFTofDE, together with built-in mappings to a number of well-known theories, including $f(R)$ gravity, quintessence, and several classes of scalar–tensor models. It provided an important complement to the benchmark code used to explore modified gravity at that time, \texttt{MGCAMB}, allowing to treat specific models without having to resort to the quasi-static approximation (see e.g.~\cite{ExtendedG,BeyondH,Horava,K-mouflage,Lin:2019qug,2015PhRvD..91f3524H}). A very important and novel feature of \eft~was the \emph{stability module} which enabled the check for physical viability of any chosen model. This has shown powerful in constraining the parameter space prior to fit to the data  and enabling physically informed reconstruction of gravity and dark energy from cosmological data sets~\cite{Raveri:2017qvt,Peirone:2017ywi,Espejo:2018hxa,Frusciante:2018vht,Pogosian:2021mcs,Raveri:2021dbu,deBoe:2024gpf}. These capabilities enabled a broad range of applications, from parameter constraints using CMB and LSS data to forecasts for future surveys, and established \eft~as one of the standard tools for model-independent tests of gravity. However, the original implementation was primarily designed around specific parametrizations or a limited set of mapped covariant models, and made the inclusion of new theories technically involved and less flexible. In particular, there was no general module capable of handling arbitrary covariant scalar–tensor theories directly from their Lagrangian, nor a fully non-parametric interface for reconstructing the EFT functions. In addition, earlier versions relied mainly on the native \texttt{CAMB} interface and CosmoMC, which limited their integration with modern parameter inference pipelines. As the theoretical landscape has expanded and observational precision has increased, these limitations have become more apparent, motivating a substantial upgrade of the code. The new version presented in this work addresses these issues by introducing a fully covariant Horndeski module, a non-parametric implementation of the EFT functions, and a Python wrapper with seamless integration into the \texttt{Cobaya} framework. Together, these developments significantly extend the scope, flexibility, and usability of the \eft~framework, facilitating both theory-driven model testing and large-scale data analyses.

\begin{figure}[htbp]
\centering
\hspace*{-0.15\textwidth}
\includegraphics[width=1.3\textwidth]{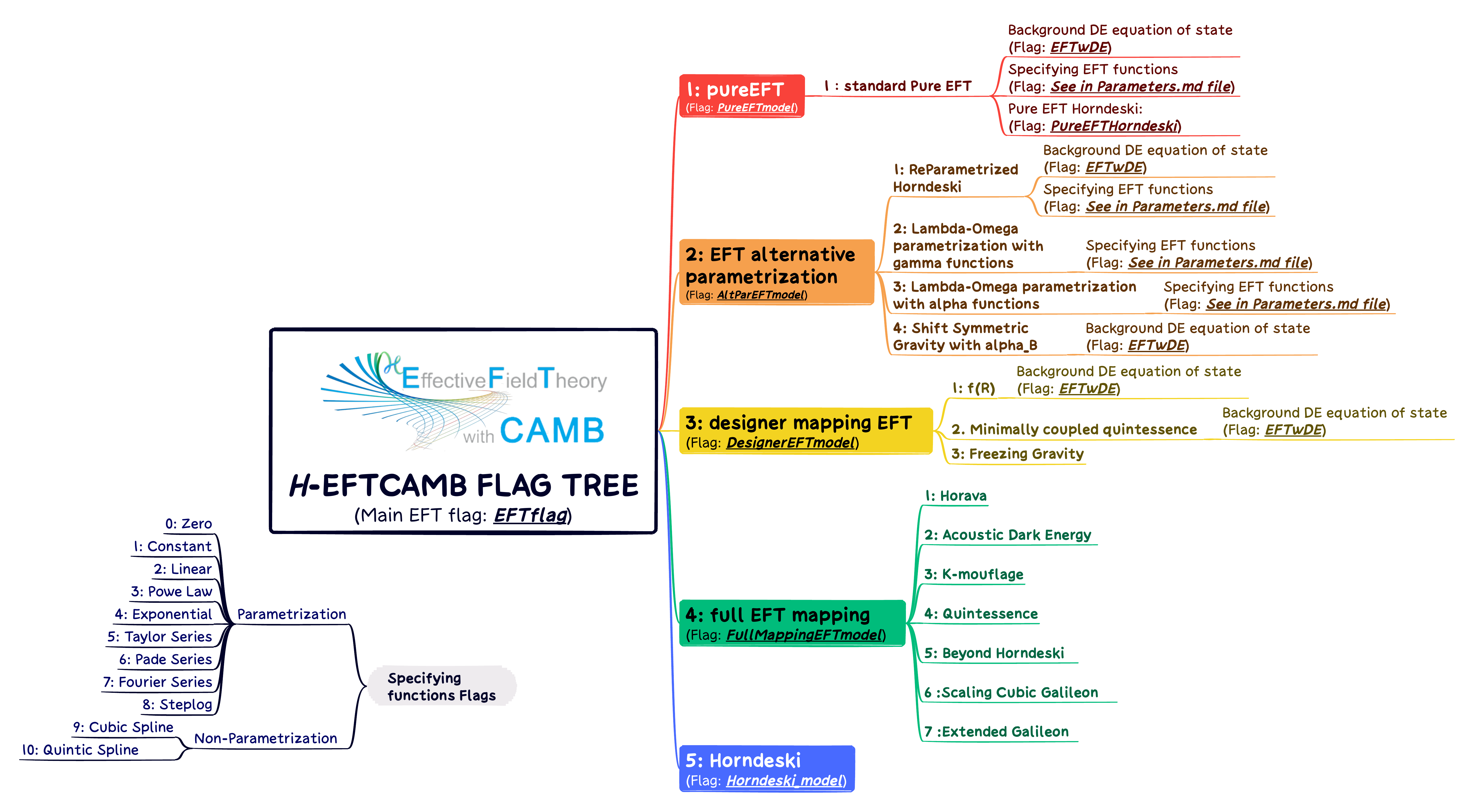}
\caption{The flag tree for \heft. The main branch tree displays the flag structure for model selection in \heft. The separate tree in the bottom left lists the flags used to specify functions.}
\label{fig:tree}
\end{figure}

\begin{figure}[htbp]
\centering
\includegraphics[width=1\textwidth]{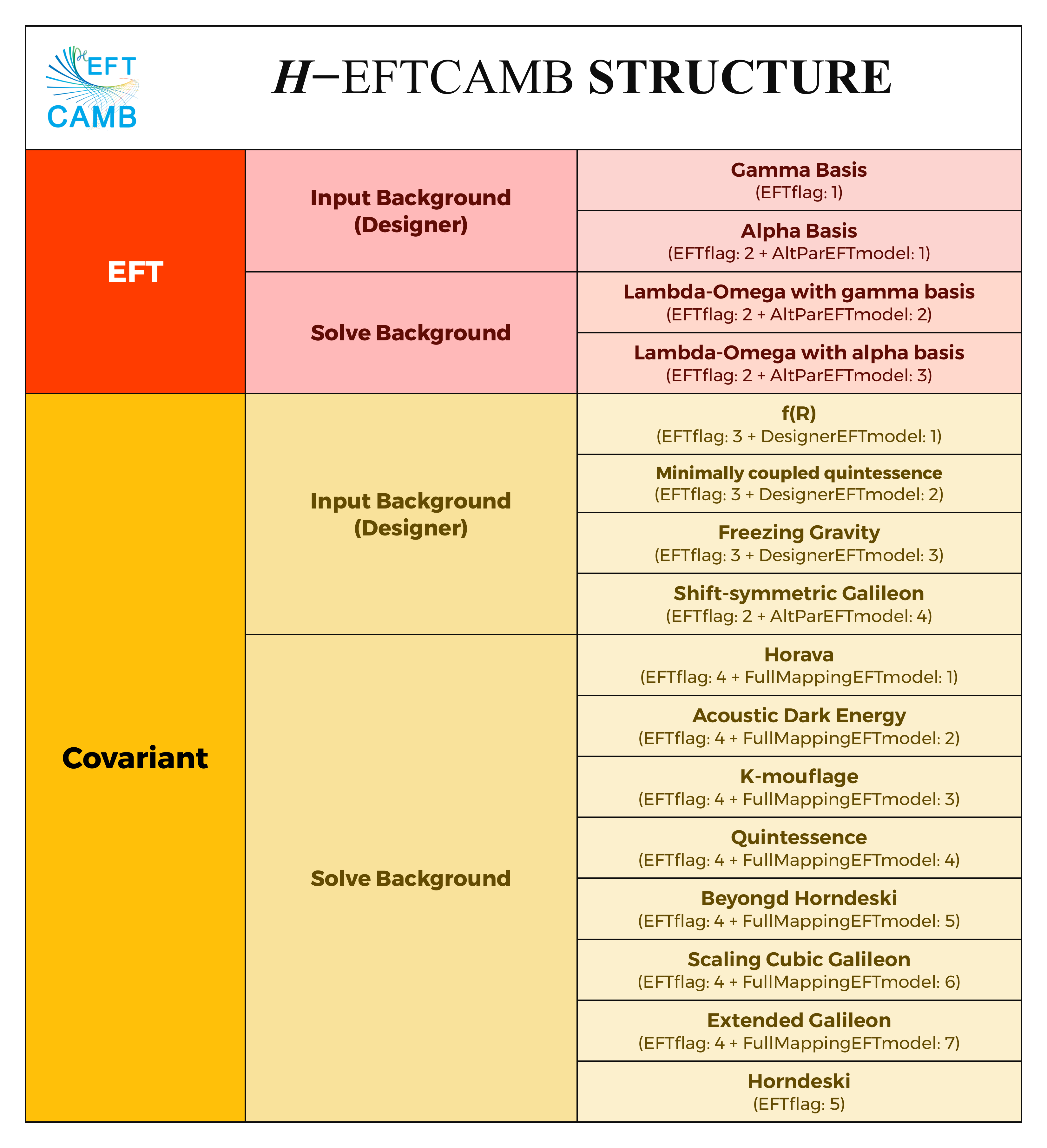}
\caption{The physical model structure of \heft. The first column refers to whether a model is defined by specifying an EFT or covariant Lagrangian. The second column indicates whether the background is input (Designer) or derived from the corresponding Larangian (Solve Background). The parenthesis in the third column indicates the flags to select the corresponding model.}
\label{fig:chart}
\end{figure}

\subsection{The original \eft} \label{subsec:overview}

\paragraph{} \heft~includes all functionalities of the original \eft. This subsection provides a brief overview of the framework and key features of the original \eft. We refer readers to \cite{EFTCAMB1,EFTCAMB2,NumericalNotes3} for more detailed description of the original code.

In the EFTofDE framework, the background and linear perturbation dynamics is fully described by the quadratic EFT action~\cite{EFTCAMB2}
\begin{align}\label{action}
  S = \int d^4x &\sqrt{-g}   \left \{ \frac{M_p^2}{2} \l[1+\Omega(\tau)\r]R+ \Lambda(\tau) - a^2c(\tau) \delta g^{00} \right. \nonumber \\
  & \left. +\frac{M_2^4 (\tau)}{2} (a^2\delta g^{00})^2 - \frac{\bar{M}_1^3 (\tau)}{2}a^2 \delta g^{00} \delta {K}\indices{^\mu_\mu}  \right. \nonumber \\ 
  & \left. - \frac{\bar{M}_2^2 (\tau)}{2} (\delta K\indices{^\mu_\mu})^2 - \frac{\bar{M}_3^2 (\tau)}{2} \delta K\indices{^\mu_\nu} \delta K\indices{^\nu_\mu}
  +\f{a^2\hat{M}^2(\tau)}{2}\delta g^{00}\delta R^{(3)} \right. \nonumber \\
  &\left. + m_2^2 (\tau) (g^{\mu \nu} + n^\mu n^\nu) \partial_\mu (a^2g^{00}) \partial_\nu(a^2 g^{00})
  \right\} + S_{m} [\chi_i ,g_{\mu \nu}],\nonumber\\
\end{align}
whose operators are constructed by perturbation to the time-time component of the metric $\delta g^{00}$, perturbation to the intrinsic Ricci curvature $\delta R^{(3)}$ and to the extrinsic curvature tensor $\delta K_\mu^\nu$ of the constant time surface. $\tau$ is the conformal time, \( a \) is the scale factor, and \( M_p \) and \( R \) denote the reduced Planck mass and the spacetime Ricci scalar, respectively. Action \eqref{action} uniquely determines the background and linear perturbation dynamics given the EFT functions \( \{ \Omega, \Lambda, c, M_2, \bar{M}_1, \bar{M}_2, \bar{M}_3, \hat{M}, m_2 \} \), which are all functions of time only. 
The basic idea of \eft~is to numerically evolve the DE dynamics based on \eqref{action} with the EFT functions as input. 

The background dynamics is described by three EFT functions $\{\Omega,\Lambda,c\}$, only two of which are dynamically independent while the other is determined through the equation of motion. By convention, in \eft~we always treat $\{\Omega,\Lambda\}$ as independent and $c$ as derived. \eft~supports two methods of specifying the background: ``Designer" and ``Solving background". In the former the user inputs one of $\{\Omega,\Lambda\}$ and the background cosmology (e.g. by inputting a $w_{\rm DE}(a)$), the code then computes the other background EFT function through the equation of motion. In the latter the user inputs the full background Lagrangian (i.e. specifying both $\{\Omega,\Lambda\}$) and the code computes the background evolution. The ``Designer" approach is particularly useful when one wants to explore DE dynamics in a known background, usually informed by observation. In contrast, the ``Solving background" approach is more theoretical as it starts from fully specifying an EFT action.

The rest six EFT functions only affect the perturbation dynamics. In \eft~they are conventionally represented by six dimensionless time-dependent parameters, dubbed the gamma basis:
\begin{align}
&\gamma_1=\frac{M^4_2}{M_p^2H_0^2}, \quad \gamma_2=\frac{\bar{M}^3_1}{M_p^2H_0}, \quad \gamma_3=\frac{\bar{M}^2_2}{M_p^2}, \nonumber \\
&\gamma_4=\frac{\bar{M}^2_3}{M_p^2}, \quad \gamma_5=\frac{\hat{M}^2}{M_p^2}, \quad \gamma_6=\frac{m^2_2}{M_p^2}
\end{align}
\eft~computes the background and linear perturbation dynamics with the input $\{\Omega,\Lambda\}$ and $\gamma_i$'s as functions of the scale factor. In particular, the Horndeski theory corresponds to
\begin{equation}\label{eq:Horndeski_condition}
2\gamma_5 = \gamma_3 = -\gamma_4,\quad \gamma_6 = 0.
\end{equation}

Another widely used parameterization of the Horndeski theory is the alpha basis, where the background and linear perturbation dynamics are fixed by $\{H, \alpha_K, \alpha_B, \alpha_T, \alpha_M\}$ where $H(a)$ is the background evolution and $\alpha_i$'s are four functions of the scale factor describing the DE/MG dynamics \cite{Bellini:2014fua}. \eft~supports this parametrization by mapping it back to the gamma basis, see \cite{NumericalNotes3} for details.

\subsection{New features of \heft}

The major new features of \heft~are summarized as:
\begin{itemize}
    \item A direct module that operates with any covariant Lagrangian within the Horndeski class.
    \item A highly flexible Python wrapper that seamlessly integrates with \texttt{Cobaya}.
    \item New parametric and non-parametric methods of defining free functions.
    \item Inclusion of several DE theories mapping to the EFT framework, new initial conditions consistent with non-zero EFT functions in the early time and the positivity bound.
\end{itemize}

To maintain compatibility with codes making use of the old \eft, \heft~retains the same flag tree structure of selecting DE/MG models, as depicted in Fig.\ref{fig:tree}. Model selection can be read out from the tree in this way: at each level the name box includes a flag name whose integer value selects the subsequent branch, e.g. the red box on the top with ``1: pureEFT" indicates that the pureEFT branch is selected by \texttt{EFTflag = 1}. For example, the \texttt{Scaling Cubic Galilleon} model can be selected by \texttt{EFTflag = 4, FullMappingEFTmodel = 6}. More and more models are added to \eft~during the past decade, making the naming and branching of the flag tree ambiguous. Therefore, we regrouped the models in Fig.~\ref{fig:chart} to provide a clearer and more physical overview of the structure of \heft. In \heft, a model is defined by either specifying an EFT or covariant action, which is indicated by the first column of Fig.\ref{fig:chart}. Generally, one also has two ways to deal with the background: specifying all parameters of the action and derive the background from it (Solve Background) or  inputting the background time evolution, e.g. $w_{\rm DE}(a)$, to trade for one free function in the Lagrangian through equation of motion (Designer), which is represented by the second column of Fig.\ref{fig:chart}. The third column of Fig.\ref{fig:chart} reports the supported models and the corresponding selection flags.

\subsubsection{The Horndeski module}\label{subsec:Horndeski module}

The Horndeski module is based on the most general covariant Horndeski action
\begin{align}
    \mathcal{S}_{\mathcal{H}} = \int d^4x \sqrt{-g} \left( L_2 + L_3 + L_4 + L_5 \right),
\end{align}
where the Lagrangians have the following structure:
\begin{align}
L_2 &= G_2(\phi, X), \nonumber \\
L_3 &= G_3(\phi, X)\Box\phi, \nonumber \\
L_4 &= G_4(\phi, X)R - 2G_{4X}(\phi, X)\left[ (\Box\phi)^2 - \phi_{;\mu\nu}\phi^{;\mu\nu} \right], \nonumber \\
L_5 &= G_5(\phi, X)G_{\mu\nu}\phi^{;\mu\nu} + \frac{1}{3}G_{5X}(\phi, X) \left[ (\Box\phi)^3 - 3\Box\phi \, \phi^{;\mu\nu} \phi_{;\mu\nu} + 2\phi_{;\mu}^{\;\;\nu} \phi_{;\nu}^{\;\;\sigma} \phi_{;\sigma}^{\;\;\mu} \right].
\end{align}
Here, $G_{i}(i=2,3,4,5)$ are arbitrary functions of the scaler field $\phi$ and its kinetic term $X\equiv-\frac{1}{2}g^{\mu\
\nu}\nabla_\mu\phi\nabla_
\nu\phi$, and $G_{iX}\equiv\partial G_{i}/\partial X$. $R$ is the Ricci scalar, and $G_{\mu\nu}$ is the Einstein tensor.

For an arbitrary DE/MG theory belonging to the Horndeski class, the new Horndeski module (\texttt{EFTflag:5}) of \heft~expects its covariant Lagrangian as input, which then automatically evolves the corresponding background and linear perturbation cosmology. To solve for the scalar field perturbation, \heft~provides two options controlled by the flag \texttt{EFTCAMB\_evolve\_delta\_phi}. When \texttt{EFTCAMB\_evolve\_delta\_phi = False}, \heft~will map the covariant Horndeski theory to the corresponding EFT and solve the perturbation using the EFT framework already implemented in the original \eft. When \texttt{EFTCAMB\_evolve\_delta\_phi = True}, the code will directly integrate the equations of motions derived from the covariant Lagrangian (i.e. in terms of $\delta\phi$) which is newly implemented by \heft, see Appendix.\ref{apdx:deltaphi eq}. The latter is especially useful when the scalar field has turn overs (i.e. $\dot{\phi}$ crossing zero), e.g. oscillating fields, which are coordinate singularities in the unitary gauge and \textit{cannot} be evolved in the EFT framework. Because of this, \heft~is \textit{the only} cosmological code that can deal with general oscillating fields in the Horndeski class.

In the Horndeski module, the covariant Larangians are implemented in a single subroutine \texttt{EFTCAMBHdskLagrangianCoefficients} located in the source file \texttt{008p0\_Horndeski.f90}. This subroutine defines the Horndeski functions and their derivatives, e.g. $G_i, \partial_\phi G_i,\partial_X G_i$, as functions of any input $\phi, X$. This is also the place where any user-defined new Lagrangians should go. Models defined in \texttt{EFTCAMBHdskLagrangianCoefficients} are selected by an integer flag \texttt{Horndeski\_model}. There are some pre-implemented models, such as quintessence with power-law (\texttt{Horndeski\_model:1}) or axion-like (\texttt{Horndeski\_model:2}) potentials. Among the pre-implemented models there is a highly flexible parametrization of general Horndeski theories, dubbed \textit{freely parameterized Horndeski} (\texttt{EFTflag:5}, \texttt{Horndeski\_model:7}):
\begin{equation}\label{eq:free horndeski}
\begin{aligned}
    \mathcal{K}(\phi,X) =& sX-c_0(\phi)+c_1(\phi)X^2,\\
    G_3(\phi,X) =& c_2(\phi)X,\\
    G_4(\phi,X) =& M_p^2\left[\frac{1}{2}+c_3(\phi)+c_4(\phi)X\right],\\
    G_5(\phi,X) =& c_5(\phi)+c_6(\phi)X, 
\end{aligned}
\end{equation}
which is simply the Horndeski functions expanding to the leading non-trivial order in $X$. The sign of the canonical kinetic term $s=\pm1$ can be selected by \texttt{Horndeski\_kinetic\_coeff} and $\{c_i\}$ are functions of $\phi$ which can be freely specified by the user, either as analytic functions, function basis expansions or spline interpolations, see Fig.\ref{fig:tree} and section~\ref{subsec:free function}.

\subsubsection{The python wrapper and integration with \texttt{Cobaya}}

\heft~comes with a powerful python wrapper which exposes the main functionalities of the Fortran backend to be called in python scripts as well as modern cosmological samplers such as \texttt{Cobaya}. The python wrapper can also output many intermediate results, such as background, perturbation variables and source functions, that might be needed for plotting, debugging and theoretical analysis, see the examples in the \heft~GitHub repository for more detailed documentation.

\heft~inherits the same python API of setting up a model and computing cosmology as \texttt{CAMB}. Therefore, using \heft~in \texttt{Cobaya} is as easy as putting the same \texttt{camb} theory block in the yaml file as one does with the original \texttt{CAMB} and then adding a \texttt{path} line in the \texttt{camb} block pointing to the root directory of \heft. All \heft~configuration flags and fixed parameters which are not varied in the sampling can be put into the \texttt{extra\_args} of the \texttt{camb} block. The other parameters can either be varied in the \texttt{param} block or be provided by some other theory module of \texttt{Cobaya}. One thing to note is that the python wrapper will \textit{not} tell \texttt{Cobaya} which parameters \heft~needs, thus it is the user's responsibility to specify, by adding a \texttt{require} line in the \texttt{camb} block, \textit{all} the parameters that need to be passed to \heft, except for those already present in the \texttt{extra\_args} block. Again we refer the readers to the documentation in the \heft~GitHub repository for details.

\subsubsection{Specifying free functions in \heft}\label{subsec:free function}
The original \eft~provides several predefined analytic forms and function expansions to define one-variable functions, such as the EFT functions and $w_{\rm DE}$. \heft~further expands the inventory with two spline interpolation methods (cubic and quintic), making it possible for the user to define arbitrary one-variable function by providing a dense grid of function values and the corresponding coordinates. This new functionality is extremely useful in non-parametric reconstruction where any reconstructed function can be generated by some external module and passed to \heft~as a dense grid of values. 

In \heft~this functionality is not limited to defining function of the scale factor. In the new Horndeski module, it also supports defining function of $\phi$ in the covariant Lagrangian. Furthermore, in the two designer EFT modules (\texttt{EFTflag = 1} and \texttt{EFTflag = 2, AltParEFTmodel = 1}), the EFT functions can now be defined as both functions of the scale factor and the DE energy fraction $\Omega_{\rm DE}=\rho_{\rm DE}/3M_p^2H^2$, i.e.
\begin{equation}\label{eq:function_form}
    F = f(a) + g(\Omega_{\rm DE})
\end{equation}
where $f$ and $g$ are two free functions. In particular, this includes the commonly used parametrization for the alpha basis $\alpha_i=c_i\Omega_{\rm DE}$ as a special case.

All the details of how to specify functions in \heft~together with model flag selection can be found in the \heft~GitHub repository under \texttt{find\_your\_model} folder.

\subsubsection{Built-in covariant DE models, initial conditions and the positivity bound}
During the last decade of application, several covariant DE/MG models, namely shift-symmetric Galileons~\cite{ShiftS}, K-mouflage~\cite{K-mouflage}, acoustic dark energy~\cite{Lin:2019qug}, minimally coupled Quintessence~\cite{EFTquintessence}, beyond Horndeski~\cite{BeyondH}, scaling cubic Galileon~\cite{ScalingCG}, extended Galileon~\cite{ExtendedG} and freezing Gravity~\cite{FreezingG}, have been studied in the literature and integrated into \eft~by mapping them to the corresponding EFTs. These are also released in \heft~ and can be selected by the corresponding flags as summarized in Fig.\ref{fig:tree} and Fig.\ref{fig:chart}. For brevity, we only list the models here and refer the readers to the corresponding research papers for details. Besides, the cosmological positivity bound postulated in Ref.\cite{deBoe:2024gpf} and the initial conditions consistent with non-zero EFT functions in the early times from Ref.\cite{Pan:2025upl} are also included in \heft, which can be activated by the flag \texttt{EFT\_positivity\_bounds} and \texttt{EFT\_IC\_type} respectively.

\section{Validation of \heft} \label{sec:Compare}

In this section, we present several examples demonstrating the new features of \heft~and validate their results by either internally comparing with existing results from the original \eft~or cross-checking with external code (\texttt{hi\_class}) that supports the same model. All models considered in this section are included as examples in the \heft~repository.

\subsection{Internal Comparison with EFT framework: Scaling Cubic Galileon (SCG)}\label{subsec:scg}

The implementation of SCG model into \eft~(\texttt{EFTflag:4}, \texttt{FullMappingEFTmodel:6}) was introduced in Ref.~\cite{ScalingCG}, with its Lagrangian reads:
\begin{equation}
    L_{SCG} = X - V_{1}e^{-\beta_{1}\phi} - V_{2}e^{-\beta_{2}\phi}-A\ln Y\square\phi,\ Y=Xe^{\lambda\phi}.
\end{equation}
Here, $V_1$, $V_2$, $\beta_1$, $\beta_2$, $A$, and $\lambda$ are constants, while the parameters $\{\beta_1, \beta_2, A, \lambda\}$ are treated as free parameters in this model. Under the Horndeski class, the Langrangians can be mapped to:
\begin{equation}\label{eq:SCG-HornL}
\begin{aligned}
    \mathcal{K}(\phi,X) =& X - V_{1}e^{-\beta_{1}\phi} - V_{2}e^{-\beta_{2}\phi},\\
    G_3(\phi,X) =& A\ln Y,\\
    G_4(\phi,X) =& 0,~ G_5(\phi,X) = 0.
\end{aligned}
\end{equation}
which is implemented in the Horndeski module with (\texttt{EFTflag:5}, \texttt{Horndeski\_model:6}).

The original SCG module implemented a custom set of equations specific to this model, in terms of the dynamical variables \cite{ScalingCG}
\begin{equation}
    x=\frac{\dot{\phi}}{\sqrt{6}H}, \quad y_1= \frac{\sqrt{V_1e^{-\beta_1\phi}}}{\sqrt{3}H},\quad y_2= \frac{\sqrt{V_2e^{-\beta_2\phi}}}{\sqrt{3}H}
\end{equation}
and the initial condition is provided with respect $\{x,y_1,y_2\}$. In order to compare results of the Horndeski module with the original SCG, one thus has to make sure to provide equivalent initial conditions in terms of $\{\phi_{\rm ini}, \dot{\phi}_{\rm ini}, V_1, V_2\}$. We set $\phi_{\rm ini}=0$ since difference in $\phi_{\rm ini}$ is equivalent to a constant shift in $V_1$ and $V_2$. Following \cite{ScalingCG}, $\dot{\phi}$ and $V_1$ are determined by starting from the initial tracking solution in radiation dominance, and $V_2$ is determined by a shooting algorithm enforcing the Friedman constraint today.

For comparison, we consider four sets of parameters listed in Table~\ref{tab:SCG}. Fig.\ref{sec:Compare} shows the CMB temperature and polarization spectra, as well as the matter power spectra, computed by \heft~and the SCG model in \eft. The comparison results show that the predictions from \heft~agree perfectly with those from the SCG model.

\begin{table}[h!]
\centering
\begin{tabular}{|c|c|c|c|c|c|}
\hline
Model & $\beta_1$ & $\beta_2$ & $A$ & $\lambda$\\ \hline
M1 & 100 & 0.7 & -0.3 & 154 \\ 
M2 & 100 & 0.7 & 0.09 & -8 \\ 
M3 & 100 & 0.7 & -0.28 & 148.3 \\ 
M4 & 100 & 2.5 & -1 & 150 \\ \hline
\end{tabular}
\caption{The four selected parameter sets $\{\beta_1, \beta_2, A, \lambda\}$ used for comparison.
\label{tab:SCG}}
\end{table}

\begin{figure}[htbp]
\hspace*{-0.1\textwidth}
\centering
\includegraphics[width=1.2\textwidth]{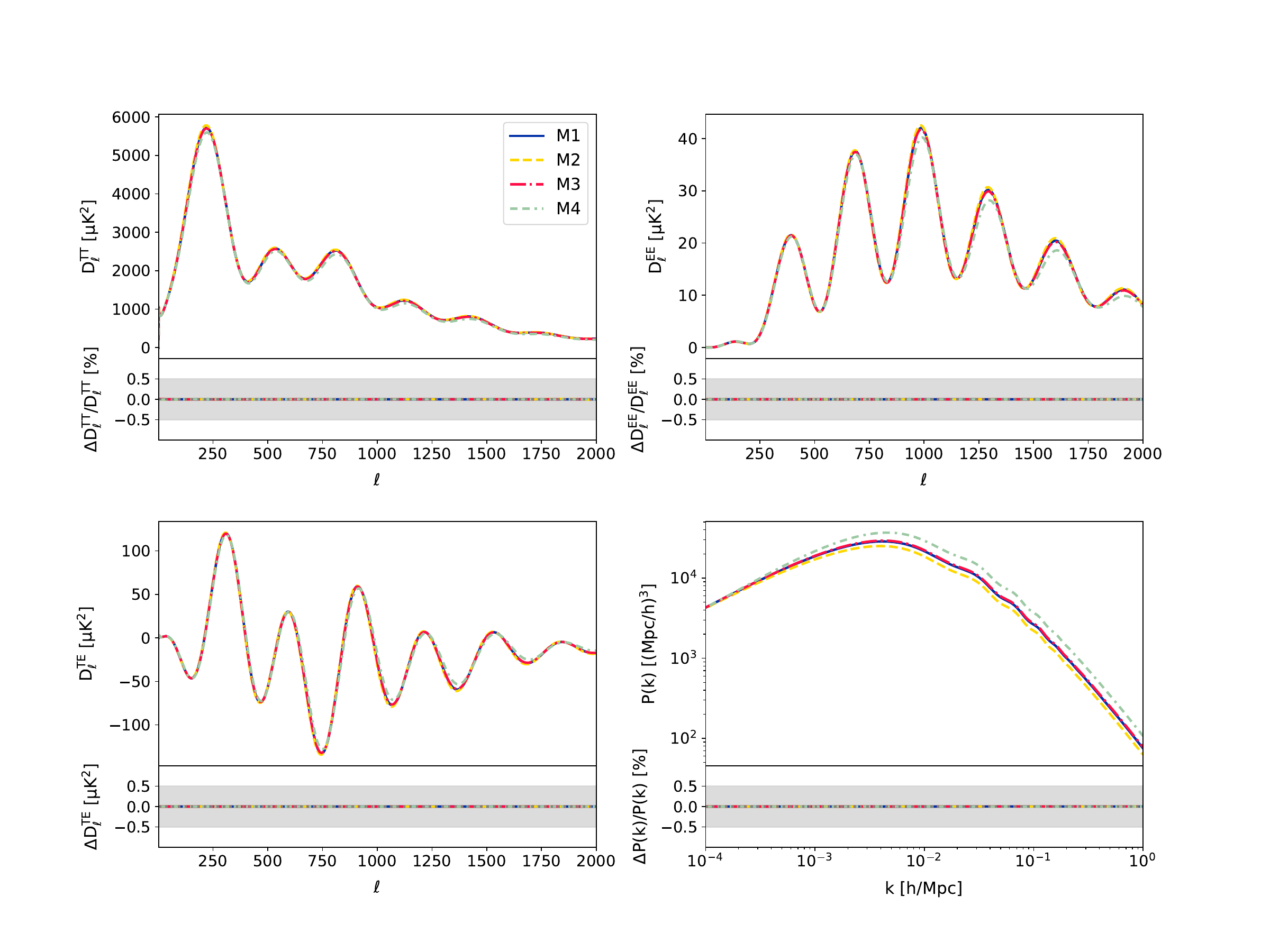}
\caption{The TT, EE, and TE angular power spectra of the CMB, together with the matter power spectrum at $z=0$ for four different sets of SCG model parameters along with the relative difference between the result from the new Horndeski module in \heft~and that from the original implementation of SCG in \eft.
\label{fig:Compare_SCG}}
\end{figure}

\subsection{External Comparison with \texttt{hi\_class}: Jordan-Brans-Dicke (JBD) theory}\label{subsec:jbd}
\begin{figure}[htbp]
\hspace*{-0.1\textwidth}
\centering
\includegraphics[width=1.2\textwidth]{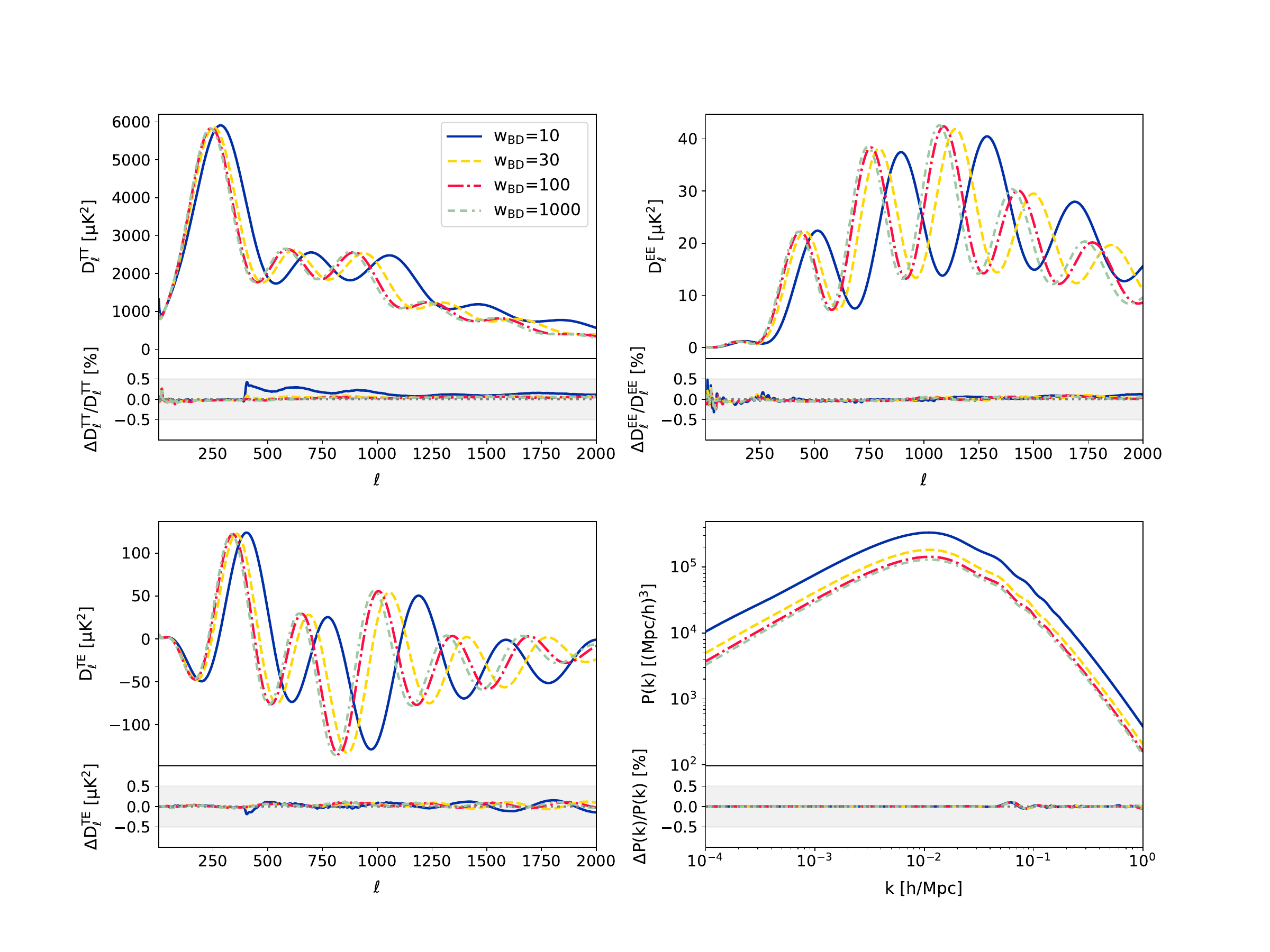}
\caption{Similar to Fig.~\ref{fig:Compare_SCG}. Comparison of the results of the JBD model with $\omega_{\rm bd} = \{10,30,100,1000\}$ from \heft~and \texttt{hi\_class}.}
\label{fig:Compare-JBD}
\end{figure}

We consider the JBD theory with a constant potential (cosmological constant)
\begin{equation}
    L_{JBD} = \frac{M_p^2}{2} \phi R - \frac{\omega_{\rm BD} M_p^2}{2\phi}(\partial\phi)^2-\Lambda.
\end{equation}
where $\omega_{\rm BD}$ is the only free parameters in this model, with the higher value indicate the lower modification of the gravity. This model is predefined in \texttt{hi\_class}. In \heft~ we use parametrization Eq.\eqref{eq:free horndeski} ((\texttt{EFTflag:5}, \texttt{Horndeski\_model:7})) to represent JBD. To this end, one has to perform a field redefinition,
\begin{equation}
    \varphi = 2\sqrt{w_{\rm BD}\phi}
\end{equation}
which transforms the kinetic term into the canonical form and the resulting Lagrangian is
\begin{equation}
    L_{JBD} = \frac{\varphi^2}{8w_{\rm BD}}R-\frac{1}{2}(\partial\varphi)^2-\Lambda
\end{equation}
corresponding to
\begin{equation}
    c_0 = \Lambda, \qquad c_3(\varphi) = -\frac{1}{2}+\frac{\varphi^2}{8\omega_{\rm BD}}
\end{equation}
in the \textit{freely parameterized Horndeski} \eqref{eq:free horndeski}.
For comparison, we fix the cosmological parameters to fiducial values and the initial field value $\phi_{\rm ini}=1$ ($\varphi=(4\omega_{\rm BD}\phi_{\rm ini})^{-1}$). In both models, $\Lambda$ is determined by a shooting algorithm that requires the calculated $H_0$ to match the input value. In \texttt{hi\_class} the shooting is hard coded for JBD. In \heft~this is implemented as an external python script which calls \heft~python wrapper. We recommend this method for the general use case of \heft~as it benefits from the great flexibility of python and the existing root finding algorithms. Figure.\ref{fig:Compare-JBD} shows the CMB temperature and polarization spectrum as well as the matter power spectrum calculated by \heft~(both $\delta\phi$ and the EFT framework) and \texttt{hi\_class} with different $\omega_{\rm BD}$. It shows that the new Horndeski module and $\delta\phi$ framework are consistent with existing codes at sub-percent precision, while enjoying much higher theoretical flexibility.

\subsection{External Comparison with \texttt{hi\_class}: $\alpha_i=c_i\Omega_{\rm DE}$ parametrization} \label{subsec:alphaDE}
A commonly studied parametrization of DE/MG is to parameterize the alpha basis as
\begin{equation}\label{eq:alphaDE}
    \alpha_i = c_i \Omega_{\rm DE}, \qquad i=K,B,T,M
\end{equation}
where $\Omega_{\rm DE}\equiv \rho_{\rm DE}/3M_p^2H^2$ is the DE energy fraction. The original \eft~does not support \eqref{eq:alphaDE} since the EFT functions can only be specified as functions of the scale factor. In the new \heft, the extended function specification described in Section.\ref{subsec:free function} now includes parameterization \eqref{eq:alphaDE} as a special case in the alpha basis model (\texttt{EFTflag = 2, AltParEFTmodel = 1}). For example, $\alpha_{M}=c_M\Omega_{\rm DE}$ corresponds to \texttt{RPHusealphaM = True, RPHalphaMmodel = 0, RPHalphaMmodel\_ODE = 2, RPHalphaM\_ODE0 = $c_M$}, where \texttt{RPHusealphaM = True} tells the code we are parameterizing $\alpha_{\rm M}\equiv \frac{d\log (M^2_*/M_p^2)}{d\ln a}$ instead of the effective Planck mass squared $M^2_*$, \texttt{RPHalphaMmodel = 0, RPHalphaMmodel\_ODE = 2} select zero for the $a$ dependent part and linear function for the $\Omega_{\rm DE}$ dependent part of $\alpha_M$ respectively (i.e. the $f$ and $g$ in Eq.\eqref{eq:function_form}, see Fig.\ref{fig:tree} for all options).

\begin{figure}[htbp]
\hspace*{-0.1\textwidth}
\centering
\includegraphics[width=1.2\textwidth]{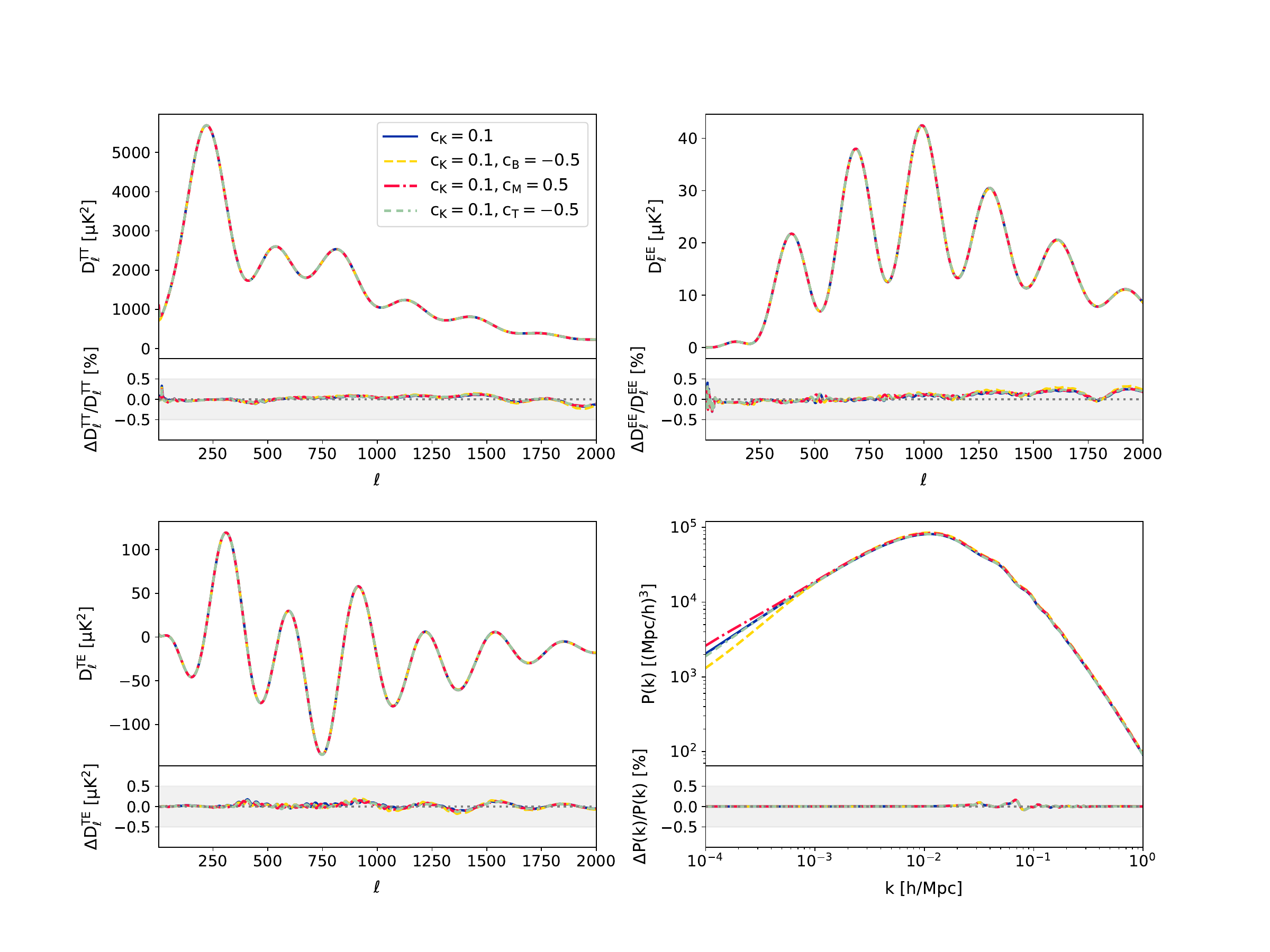}
\caption{Similar to Fig.~\ref{fig:Compare_SCG}. Comparison of the results of the $\alpha_i=c_i\Omega_{\rm DE}$ parametrization from \heft~and \texttt{hi\_class}.}
\label{fig:Compare-alphaDE}
\end{figure}

To compare with \texttt{hi\_class}, we fix the background to $\Lambda$CDM and consider three models where each one of $\{c_B,c_T,c_M\}$ is set to non-zero while the others to zero \footnote{Note that the $\alpha_B$ defined in \eft~and \heft~has a constant $-1/2$ factor difference from the one used in \texttt{hi\_class}, i.e. $(\alpha_B)_{\eft}=-\frac{1}{2}(\alpha_B)_{\texttt{hi\_class}}$.}. In all three models we fix $c_K=0.1$ since kineticity is not constrained by observation. For completeness we also include the fourth model with $c_K=0.1$ and $c_B=c_T=c_M=0$. Figure.\ref{fig:Compare-alphaDE} demonstrates the relative variance in the computed CMB spectra and matter power spectrum, proving that the results are consistent. We note that the flexibility of \heft~allows more complicated parametrization, e.g. the $\Omega_{\rm DE}$ polynomial considered in \cite{hqwq-m19h}. We choose the linear form in order to compare with \texttt{hi\_class}.

\section{Testing Gravity with \heft: An example} \label{sec:tgs}
To showcase the application of \heft~in testing gravity, we consider a thawing gravity theory with screening mechanism ($\rm TG_s$) with Lagrangian
\begin{equation}\label{eq:tgs}
    L_{TG_s} = \frac{M_p^2}{2}[1-\xi(\phi/M_p)^2]R + X + \zeta X\Box\phi - \Lambda e^{-\lambda\phi/M_p},
\end{equation}
where $\{\xi, \zeta, \lambda, \Lambda\}$ are free parameters. The above Lagrangian adds an additional Galilleon operator $X\Box\phi$ to the original thawing gravity (TG) model proposed in \cite{NMG-DR1}. Ref.~\cite{Ye:2024zpk} found such an operator might be able to screen the modified gravity effect in TG which is originally unscreened and violates solar system constraints. It is argued that the addition of this operator will not break TG's ability to explain observation and support stable phantom crossing if it is only excited at scales much smaller than the Hubble scale. To verify this argument rigorously and illustrate the use of code, we use \heft~to study the model Eq.\eqref{eq:tgs} with the coupling strength $\zeta$ of the $X\Box\phi$ operator in the range $0<\frac{\zeta}{M_p (kpc)^{-2}}<1$, which is expected to become relevant only on scales ($\sim kpc$) much smaller than current Hubble scale. In \heft~, $\rm TG_s$ is a subclass of the \textit{freely parameterized Horndeski} model~\eqref{eq:free horndeski} with
\begin{equation}
    c_0(\phi) = \Lambda e^{-\lambda\phi/M_p}, \qquad c_2=\zeta, \qquad c_3(\phi) = -\xi(\phi/M_p)^2.
\end{equation}
We performed nested sampling analysis using \texttt{Cobaya} interfaced with \texttt{Polychord} \cite{Handley:2015vkr,Handley:2015fda} over the combined dataset of Planck CMB low and high-$\ell$ TTTEEE spectra \cite{Rosenberg:2022sdy, Planck:2019nip}, Planck PR4 lensing \cite{Carron:2022eyg}, type Ia supernova from Pantheon+ \cite{Scolnic:2021amr} and DESI DR2 BAO \cite{DESI:2025zgx}. 

\begin{figure}[htbp]
\centering
\includegraphics[width=1\textwidth]{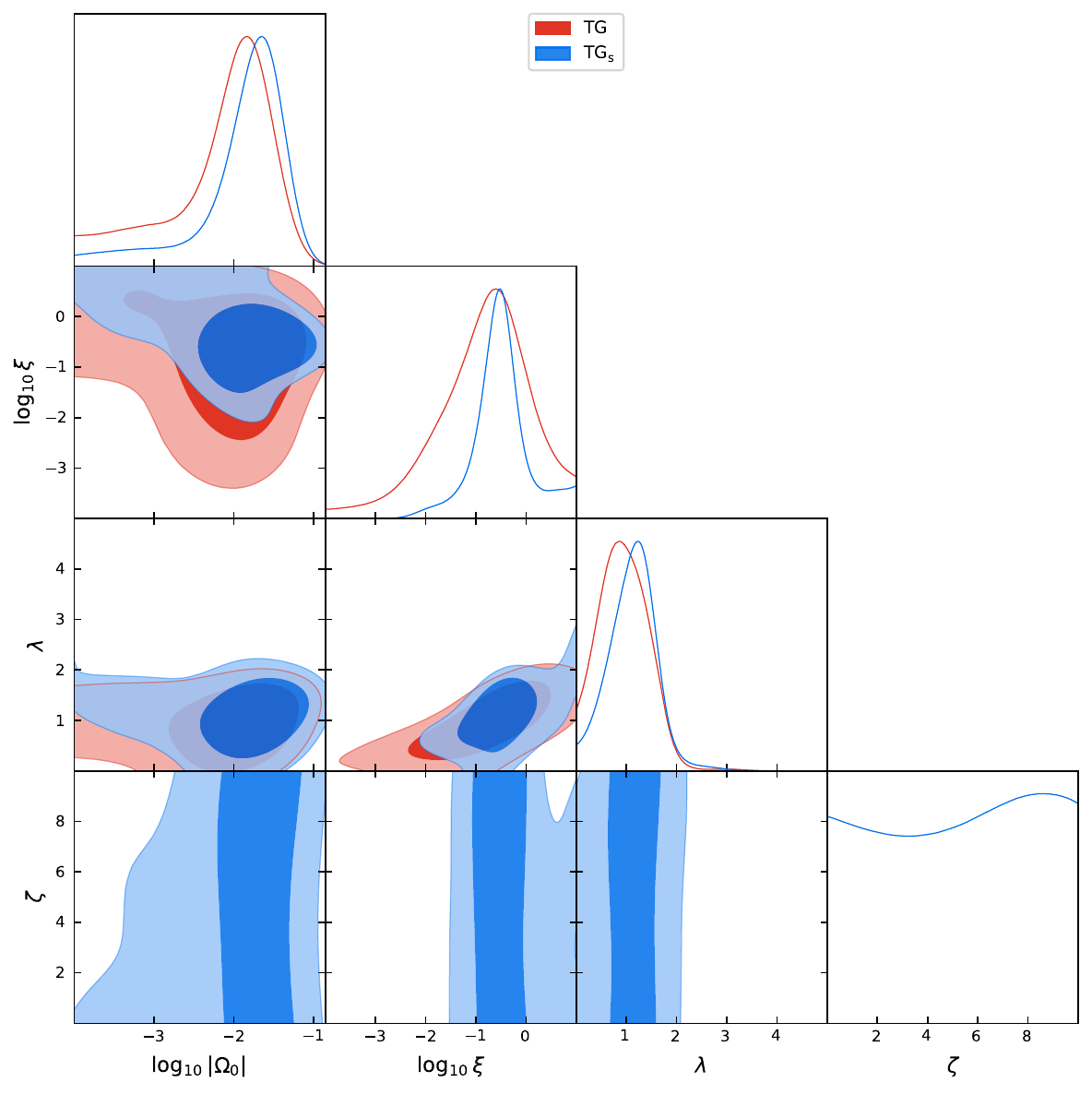}
\caption{The $1\sigma$ and $2\sigma$ posterior distributions of the model parameters in TG and $\rm TG_s$.}
\label{fig:trigTG}
\end{figure}

\begin{figure}[htbp]
\centering
\includegraphics[width=1\textwidth]{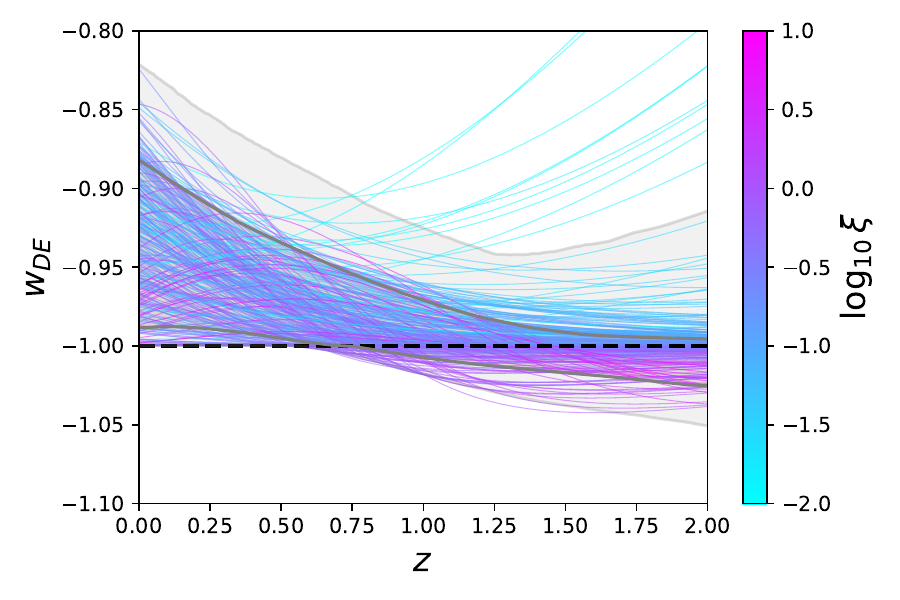}
\caption{The effective dark energy equation of state $w_{\rm DE}$ in $\rm TG_s$ for the sampled models, with a color coding for the corresponding value of non-minimal coupling parameter $\xi$ for each line. The gray shades indicate the $68\%$ and $95\%$ confidence interval of $w_{\rm DE}$.}
\label{fig:wde}
\end{figure}

Among the four theory parameters $\{\xi, \zeta, \lambda, \Lambda\}$, the energy scale $\Lambda$ of the scalar potential is not free. It can be determined by the Friedman equation similar to the $\Lambda$ in $\Lambda$CDM. In addition to $\{\xi, \zeta, \lambda\}$, there is one more free parameter describing the initial condition of the scalar field $\phi_{\rm ini}$, in place of which we vary instead $\Omega_{0}\equiv -\xi(\phi_{\rm ini}/M_p)^2$ and set $\dot{\phi}_{\rm ini}=0$ \footnote{Given the existence of the operator $X\Box\phi$, one should also start from a non-vanishing $\dot{\phi}_{\rm ini}$. In practice, we notice that the numerical evolution with $\dot{\phi}_{\rm ini}=0$ quickly relaxes to the consistent solution deep in the radiation dominance thus should not impact the data analysis. Given the main scope of this paper is to present the new \heft, we choose to set $\dot{\phi}_{\rm ini}=0$ and leave more detailed theoretical consideration to future studies.}. In total we vary the six cosmological parameters $\{\Omega_ch^2,\Omega_bh^2,H_0,\ln 10^{10}A_s,n_s\}$ together with the four model parameters $\{\lambda, \zeta, \log_{10}\xi, \log_{10}\Omega_0\}$. Fig.\ref{fig:trigTG} shows the posterior distribution of the $\rm TG_s$ model parameters, compared previous results of TG from \cite{Ye:2024zpk}. Fig.\ref{fig:wde} plots the effective dark energy equation of state $w_{\rm DE}\equiv\frac{-2\dot{H}-3H^2-P_m}{3H^2-\rho_m}$ for each models in the thinned samples from the data analysis, with color indicates the corresponding value of the non-minimal coupling parameter $\xi$. As expected, the analysis confirms that the new $X\Box\phi$ operator does not break the theory's ability to explain current data and stably support phantom crossing. Note that when $\xi>3/16$ and $\phi_{\rm ini}$ the TG field is oscillating near recombination \cite{Ye:2024zpk}, thus cannot be handled by \texttt{hi\_class}.

\section{Conclusion}

In this paper, we have presented a major update of \eft, originally designed as a consistent and numerically stable implementation of the EFTofDE. The new version, renamed to be \heft, significantly extended the scope and functionality of the original \eft. The main highlights of its new features  are:
\begin{itemize}
    \item A direct module that operates with any covariant Lagrangian within the Horndeski class.
    \item A highly flexible Python wrapper that seamlessly integrates with \texttt{Cobaya}.
    \item New parametric and non-parametric methods of defining free functions.
    \item Inclusion of several DE theories mapping to the EFT framework, new initial conditions consistent with non-zero EFT functions in the early time and the positivity bound.
\end{itemize}

\heft~is the official successor to \eft~and all subsequent support and development will happen under the name of \heft. A consistent quasi-static approximation (QSA) scheme is almost ready and is planned for the next update. In the long term our major focus of the future development is a consistent nonlinear scheme for general dark energy and modified gravity models, aiming to unlock the full power of upcoming cosmological surveys in testing gravity and dark energy. 

In the current version, some DE/MG models can encounter inaccuracy under default precision settings when one sets the \texttt{CAMB} precision parameter \texttt{lens\_potential\_accuracy > 1}, mainly due to numeric inaccuracy in evolving scalar field perturbations on small scales where rapid oscillation is present. To this end, without QSA, we recommend setting \texttt{lens\_potential\allowbreak\_accuracy = 0} and \texttt{nonlinear = False}. Otherwise, precision parameters, particular those relevant to perturbation integration precision and modes sampling, might need to be tuned accordingly.

\acknowledgments
We acknowledge all original developers of \eft~and everyone who has contributed to the developement of the code. GY developed the Horndeski module and the new parametrizations of \heft, and made major contribution to the new positivity bounds and Cobaya integration. SJL performed all the validation of the new code, run the chains, prepared all the examples and wrote the documentation. JP derived and implemented the new EFT initial conditions. DdB derived the positivity bounds. DdB and SV helped with major debug process of the new code. MR, BH, NF and AS supported and supervised the project, with MR further contributed the Cobaya integration and supervised the code publication. GY acknowledges support from the Swiss National Science Foundation. AS and DdB acknowledge support from the European Research Council under the H2020 ERC Consolidator Grant “Gravitational Physics from the Universe Large scales Evolution” (Grant No. 101126217 — GraviPULSE). BH and SJL are supported in part by the National Natural Science Foundation of China Grants No. 12333001 and by the China Manned Spaced (CMS) program with grant No. CMS-CSST-2025-A04. NF acknowledges  the Istituto Nazionale di Fisica Nucleare (INFN) Sez. di Napoli, Iniziativa Specifica InDark.
MR acknowledges the Istituto Nazionale di Fisica Nucleare (INFN) Sez. di Genova, Iniziativa Specifica InDark.
Computing resources were provided by the National Energy Research Scientific Computing Center (NERSC), a U.S. Department of Energy Office of Science User Facility operated under Contract No. DE-SC0009924, and by the University of Chicago Research Computing Center through the Kavli Institute for Cosmological Physics.

\appendix

\section{$\delta\phi$ equations} \label{apdx:deltaphi eq}
When the flag \texttt{EFTCAMB\_evolve\_delta\_phi = True}, \heft \ will solve for a new set of perturbation equations written in terms of $\delta\phi$. They are the dynamic equation of $\delta\phi$
\begin{eqnarray}
 \delta\phi'' + 2(1+s_{\phi,1})\mathcal{H}\delta\phi'+((1+s_{\phi,2})k^2+s_{\phi,3}\mathcal{H}^2)\delta\phi = s_{\phi,\eta} \frac{k^2\eta}{\mathcal{H}} + s_{\phi,h}h'+s_{\phi,m}\delta\rho_m
\end{eqnarray}
and the components of the modified Einstein equations
\begin{eqnarray}
(00)&\qquad h'=& \frac{4k^2\eta}{(2-\alpha_B)\mathcal{H}}+\frac{2a^2\delta\rho_m}{(2-\alpha_B)\mathcal{H}M_*^2}+(s_{00}+s_{00,k}k^2)\delta\phi+s_{00,p}\delta\phi'\\
(0i)&\qquad \eta' =& \frac{a^2}{2k^2M_*^2}\theta_m+s_{0i}\delta\phi+s_{0i,p}\delta\phi'\\
(ij)&\qquad\sigma' =& (1+\alpha_T)\eta-(2+\alpha_M)\mathcal{H}\sigma-\frac{a^2P_m\Pi_m}{M_*^2k^2}+s_{ij}\delta\phi\\
(ii)&\qquad h''=&-(2+\alpha_M)\mathcal{H}h'+2(1+\alpha_T)k^2\eta-\frac{3a^2}{M_*^2}\delta P_m\nonumber\\&&+(s_{ii}+s_{ii,k}k^2)\delta\phi+s_{ii,p}\delta\phi'+s_{ii,pp}\delta\phi''
\end{eqnarray}
where $\eta$ and $h$ are the synchronous gauge metric perturbations and $\sigma\equiv (h'+6\eta')/2k^2$. $\{\alpha_{M,B,T}, M_*^2\}$ are the Horndeski functions defined in \cite{Bellini:2014fua}. $\delta\rho_m\equiv -(\delta T_m)_0^0, \ \delta P_m\equiv(\delta T_m)^i_i/3, \allowbreak \ \theta_m\equiv ik_j(\delta T_{m})^0_j/(\rho_m+P_m), \ \Pi_m\equiv-3(\hat{k}^i\hat{k}_j-\delta^i_j/3)((T_m)_i^j-\delta_i^j(T_m)^l_l/3)/2P_m$ are the density, pressure, velocity divergence and anisotropic stress perturbations of total matter (everything else except for the Horndeski field) stress energy tensor $(T_m)_{\mu\nu}$. Prime denotes derivative with respect to the conformal time. When $2-\alpha_B\ne0$ in the entire evolution, the $(0i)$ equation is used to determine $h'$ algebraically. When $2-\alpha_B$ crosses zero the trace equation $(ij)$ is used instead to integrate $h'$. Unlike the EFT perturbation variable $\pi$ (Goldstone), the explicit expressions of the coefficients $\{s_{\phi,1},s_{\phi,2},s_{\phi,3},s_{\phi,\eta},s_{\phi,h},s_{\phi,m},s_{00}, s_{00,k}, s_{00,p},s_{0i},s_{s0i,p},s_{ij},s_{ii},s_{ii,k},s_{ii,p},s_{ii,pp}\}$ are extremely lengthy and complicated due to field redefinition redundancy $\phi\to\tilde{\phi}=f(\phi)$. Note they are not the same as directly substituting $\pi=\delta\phi/\phi'$ back into the EFT perturbation equations because background equations of motion must be used to analytically cancel out all $\phi'$ in the denominator to eliminate numerical singularity.


\bibliographystyle{JHEP}
\bibliography{biblio.bib}

\end{document}